\documentclass{PoS}

\def\to{\rightarrow}

\def\bi{\begin{itemize}}
\def\ei{\end{itemize}}

\def\tst{\tilde t}

\def\tg{\tilde g}

\def\alt{\lesssim}
\def\agt{\gtrsim}

\title{Naturalness and light higgsinos: \\
A powerful reason to build the ILC}

\ShortTitle{Naturalness and light higgsinos}

\author{\speaker{Howard Baer}\\
        Univesity of Oklahoma, Norman, OK 73019, USA\\
        E-mail: \email{baer@ou.edu}}

\author{\speaker{Jenny List}\\
        DESY, 22603 Hamburg, Germany\\
        E-mail: \email{jenny.list@desy.de}}

\author{Mikael Berggren, Suvi-Leena Lehtinen\\
        DESY, 22603 Hamburg, Germany\\
        E-mail: \email{mikael.berggren@desy.de, suvi-leena.lehtinen@desy.de}}

\author{Keisuke Fujii, Jacqueline Yan\\
        KEK, Tsukuba, Japan\\
        E-mail: \email{keisuke.fujii@kek.jp, jackie@post.kek.jp}}

\author{Tomohiko Tanabe\\
        International Center for Elementary Particle Physics, 
        University of Tokyo, Tokyo, Japan\\
        E-mail: \email{tomohiko@icepp.s.u-tokyo.ac.jp}}

\abstract{
A core prediction of natural Supersymmetry is the
existence of four light higgsinos not too far above the
mass of the $Z$ boson. The small mass splittings amongst
the higgsinos -- typically 5-20\,GeV -- 
imply very little visible energy release from decays of heavier
higgsinos.
In particular, if other SUSY particles are quite heavy,
as can be the case in SUSY with radiatively-driven naturalness,
the higgsinos are extremely hard to detect at hadron
colliders. The clean environment of electron-positron
colliders with $\sqrt{s} > 2m_{\mathrm{higgsino}}$, however, would allow for a decisive 
search for the required light higgsinos. 
Thus, $e^+e^-$ colliders should either discover or exclude natural SUSY.
We present a detailed study of higgsino pair production at 
the proposed International Linear $e^+e^-$  Collider which is under 
consideration for construction in Japan.
A variety of precision measurements should allow for 
extraction of underlying parameters and provide a window onto
physics at the grand unified scale.}

\FullConference{38th International Conference on High Energy Physics\\
		3-10 August 2016\\
		Chicago, USA}

\begin{document}

\section{Motivation}
\label{sec:intro}

In thinking about physics Beyond the Standard Model (BSM), especially when it comes to 
motivation for constructing costly new facilities, it pays to bear in mind two gems of wisdom:
\begin{itemize}
\item {\it Everything should be made as simple as possible, but not simpler.} (A. Einstein)
\item {\it The appearance of fine-tuning in a scientific theory is like a cry of distress from nature, 
complaining that something needs to be better explained.} (S. Weinberg)
\end{itemize}
The first of these we interpret as meaning: the further one strays from the Standard Model (SM), 
the more likely one is to be wrong. 
The second we interpret as meaning: 
{\it nature is natural} so any theory depending on an unexplained fine-tuning is likely a wrong theory.

Now, whereas the SM provides an excellent description of a vast panoply of data, it is known to be
highly unnatural in the scalar (Higgs) sector if its regime of validity as an effective theory is taken
much beyond the $\Lambda\sim 1$ TeV energy scale. In years past, this fact had led some theorists
to question whether fundamental scalar fields could exist in nature. Many solutions to the SM naturalness problem have been invoked, but only one appears consistent with data and with the above two missives:
extending the underlying spacetime symmetry group to its most general structure: the super-Poincar\'e 
algebra, or supersymmetry (SUSY)\cite{review}. 
Under SUSY, quadratic divergences neatly cancel to all orders in perturbation theory, 
rendering the theory {\it natural}. 
It may even be claimed that SUSY theories have generated three predictions found to be in accord with data: 
1. the measured relative strengths of the weak scale gauge couplings, 
2. the value of the top quark mass is just what is needed to generate a radiative breakdown of electroweak symmetry and 
3. the measured value of the new-found Higgs boson mass $m_h=125.1$ GeV\cite{atlas_h,cms_h} 
lies squarely within the narrow window predicted by the Minimal Supersymmetric Standard Model (MSSM)\cite{mhiggs}.

Alas, a fourth prediction, the existence of superpartner matter states, has yet to be verified. 
Recent searches at LHC with $\sqrt{s}=13$ TeV find, within the context of various simplified models, 
that $m_{\tg}\agt 1.9$ TeV\cite{atlas_gl} and $m_{\tst_1}\agt 0.85$ TeV\cite{atlas_t1}. 
In addition, the measured value of $m_h=125.1$ GeV seems to require highly mixed TeV-scale top-squarks{\cite{h125}. 
These various mass bounds are in deep discord from early naturalness constraints, 
this time arising from logarithmic rather than quadratic divergences, that $m_{\tg}\alt 350$ GeV\cite{Barbieri:1987fn} 
and that $m_{\tst_1}\alt 350$ GeV\cite{Dimopoulos:1995mi} (for fine-tuning measure $\Delta_{BG}<30$). 
This discord has led many physicists to question whether weak scale SUSY is indeed 
the path forward after all\cite{Lykken:2014bca}.
It has also led to renewed scrutiny as to the validity of the earlier naturalness 
estimates\cite{ltr,rns,Baer:2013gva,Baer:2014ica,mt}.

The most direct relation between the observed value of the weak scale and elements of the SUSY
Lagrangian comes from minimizing the MSSM scalar potential. Then it is found that
\begin{equation}
\frac{m_Z^2}{2}=\frac{m_{H_d}^2+\Sigma_d^d-(m_{H_u}^2+\Sigma_u^u)\tan^2\beta}{\tan^2\beta -1}-\mu^2
\simeq -m_{H_u}^2-\Sigma_u^u-\mu^2
\label{eq:mzs}
\end{equation}
where $\tan\beta\equiv v_u/v_d$ is the ratio of Higgs field vacuum expectation values, $m_{H_u}^2$ and $m_{H_d}^2$
are Higgs field soft SUSY breaking terms and $\mu$ is the superpotential Higgs/higgsino mass term.
If any term on the right-hand-side is far greater than $m_Z^2/2$, then some other 
(completely unrelated) term would have to be (implausibly) fine-tuned to an opposite-sign value such 
as to maintain $m_Z$ at its measured value. Alternatively, for a natural theory, all terms on the RHS
should be comparable to or smaller than $m_Z^2/2$. The {\it electroweak} fine-tuning measure
$\Delta_{EW}$ was introduced\cite{ltr,rns} to measure the largest contribution on the RHS of Eq. \ref{eq:mzs}
compared to $m_Z^2/2$. 
Then, for a natural ({\it i.e.} plausible) theory with low $\Delta_{EW}$, one finds the following.
1. The soft parameter $m_{H_u}^2$ is driven radiatively to small negative values at the weak scale. 2. In order that the radiative corrections $\Sigma_u^u\alt m_Z^2/2$, the top-squarks
are highly mixed (large $A_t$) and not too far beyond the few TeV range. 
This last requirement is completely consistent with the measured value of $m_h$ 
and with expectations from the measured value of $BF(b\to s\gamma )$. 
In addition, the gluino mass feeds into the $\Sigma$ terms and thus
$m_{\tg}\alt 4$ TeV (perhaps well beyond the reach of even HL-LHC)\cite{rns,upper}.
3. Finally, $\mu\sim 100-300$ GeV, the closer to $m_Z$ the better. 

This latter condition implies the existence of four higgsino-like states 
$\tilde{\chi}_1^\pm$ and $\tilde{\chi}_{1,2}^0$
with mass $\sim 100-300$ GeV (the closer to $m_Z$ the better). These are the only SUSY 
particles required to be near the weak scale. Furthermore, in spite of their light mass, they
are very difficult to detect at LHC due to their compressed spectra: the lightest higgsino-like
electroweakino (EWino) is typically just 5-20 GeV lighter than the heavier ones. 
Since $\tilde{\chi}_1^0$ would constitute a portion of the dark matter 
(along with perhaps axions\cite{Bae:2013bva}), 
the bulk of energy produced from higgsino decays becomes bound up in the $\tilde{\chi}_1^0$ mass: thus, the lightest state is totally invisible while the heavier states lead only to soft particles
which are difficult to distinguish from QCD backgrounds. 

On the other hand, these light higgsinos would be easily visible at an 
$e^+e^-$ collider such as ILC with $\sqrt{s}>2m_{\mathrm{higgsino}}$. 
The ILC is a proposed $e^+e^-$ linear collider under consideration for 
construction in Japan.
Just above kinematic threshold, the higgsino pair
production cross sections lie in the $10^{2}-10^3$ fb range. 
Such a machine, intended to be a Higgs factory, 
would turn out to be as well a higgsino factory\cite{Baer:2014yta}!

Overall, low values of $\Delta_{EW}\alt 30$\footnote{The onset of fine-tuning for $\Delta_{EW}\agt 30$ 
is visually displayed in Ref. \cite{upper}.} allow for a Little Hierarchy (LH) $\mu\ll m_{SUSY}$, 
but such a LH is not a problem since $m_{SUSY}$ enters the weak scale via radiative corrections 
while $\mu$ enters the weak scale at tree level.

What of the earlier naturalness estimates? 
The BG measure $\Delta_{BG}\equiv max_i|\frac{\partial\ln m_Z^2}{\partial p_i}|$ 
(where $i$ labels fundamental parameter $p_i$) was typically evaluated in terms of 
multi-soft-parameter effective theories. In more fundamental theories where the soft terms are 
all derived in terms of say the more fundamental gravitino mass $m_{3/2}$, 
then $\Delta_{BG}$ reduces to $\Delta_{EW}$\cite{Baer:2014ica}.
Other evaluations of logarithmic corrections to $m_{H_u}^2$ compared to $m_h^2$\cite{oldnsusy} 
were found to neglect the $m_{H_u}^2$ self-contribution to RG running\cite{Baer:2013gva}. 
By including this, one allows for {\it radiatvely-driven naturalness} (RNS) where the large top-Yukawa coupling 
and large soft terms drive $m_{H_u}^2$ from large, unnatural high scale values to natural values at the weak scale.

\section{Natural SUSY at the ILC: From masses and cross sections to SUSY parameters}
\label{sec:ilc}

The prospects for mass and cross section measurements have been studied in a full, 
{\sc Geant4}-based~\cite{Agostinelli:2002hh} simulation of the ILD detector concept proposed for the ILC,
using as an example the NUHM2 benchmark ILC1~\cite{Baer:2014yta}. 
Collision events were generated with {\sc Whizard 1.95}~\cite{Kilian:2007gr} including beamstrahlung and 
ISR, as well as hadronisation by {\sc Pythia 6.422}~\cite{bib:pythia} tuned to LEP data. So far, we studied 
the leptonic decay of the neutral higgsino $e^+e^- \to \tilde{\chi}_1^0\tilde{\chi}_2^0 \to \tilde{\chi}_1^0\tilde{\chi}_1^0 e^+e^- (\mu^+ \mu^-)$ and the semi-leptonic mode of charged
higgsino production $e^+e^- \to \tilde{\chi}_1^+\tilde{\chi}_1^- \to \tilde{\chi}_1^0\tilde{\chi}_1^0 q \bar{q}' e \nu_{e} (\mu \nu_{\mu})$. In both cases,
the maximal invariant mass of the (virtual) intermediate vector boson (IVB) mediating the decay
gives the mass splitting between the $\tilde{\chi}_2^0$ (or $\tilde{\chi}_1^{\pm}$) and the LSP.
The maximum energy of the same IVB gives the absolute masses via energy-momentum conservation, since at an $e^+e^-$ collider, 
the four-momentum of the initial state is known. Fig.~\ref{fig:mass}
shows the corresponding distributions for the di-electron channel after event selection. Combining electron and muon channels, all masses can be extracted with precisions of about $1\%$ from an
initial ILC data set corresponding to an integrated luminosity of $500$\,fb$^{-1}$ at $\sqrt{s}=500$\,GeV.
\begin{figure}[tbp]
\begin{center}
\includegraphics[width=0.425\textwidth]{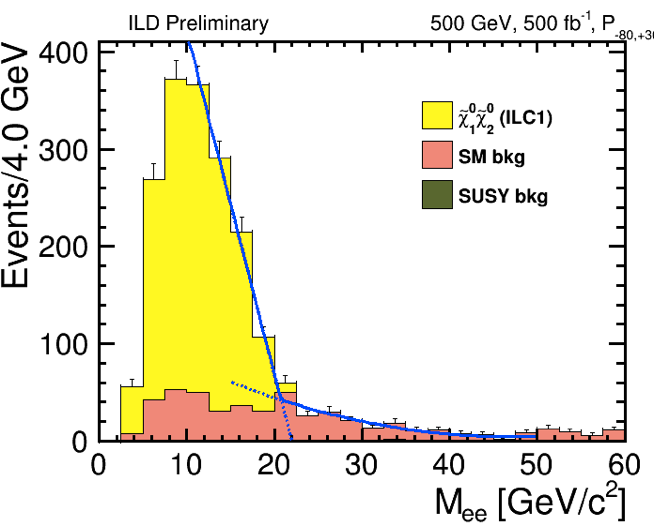}
\hspace{0.05\textwidth}
\includegraphics[width=0.475\textwidth]{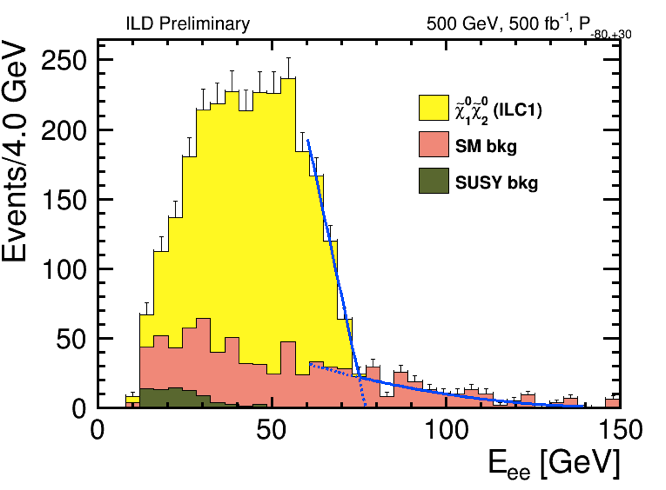}
\caption{ Neutralino mass determination in the di-electron channel: 
Left: Di-electron invariant mass. Right: Di-electron energy.
Both distributions show signal and all SM and SUSY backgrounds from full detector simulation
after event selection.
\label{fig:mass}}
\end{center}
\end{figure}

The polarised beams of the ILC allow to measure the production cross sections for 
$P(e^-,e^+) =  (+80\%,-30\%)$ and $P(e^-,e^+) =  (-80\%,+30\%)$ separately, which
reveals the higgsino nature of the charginos, as illustrated in the left panel of Fig.~\ref{fig:polxsec}.
Precisions of $3\%$ are predicted for each polarisation, again based on the initial ILC data set.
\begin{figure}[tbp]
\begin{center}
\includegraphics[width=0.425\textwidth]{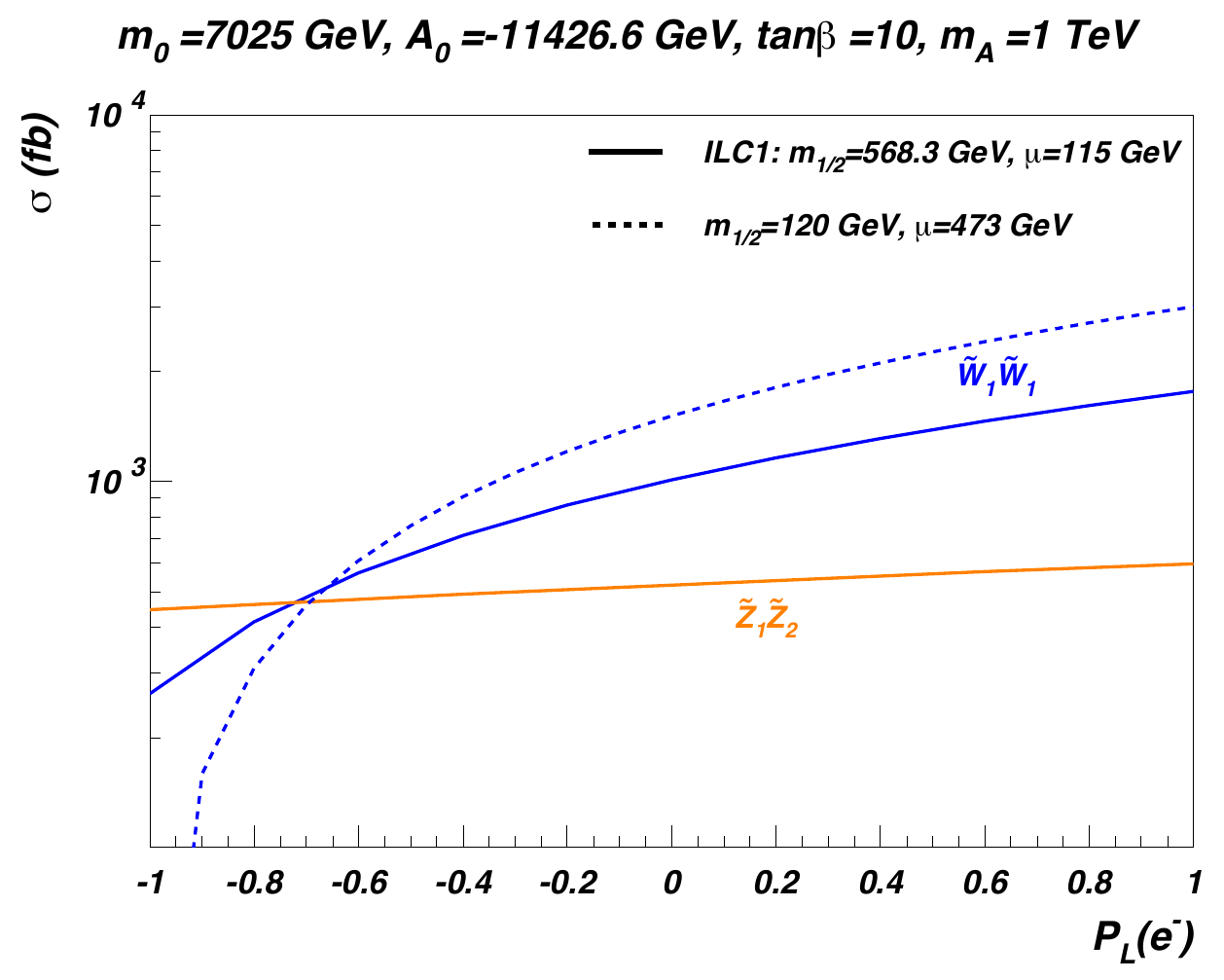}
\hspace{0.05\textwidth}
\includegraphics[width=0.475\textwidth]{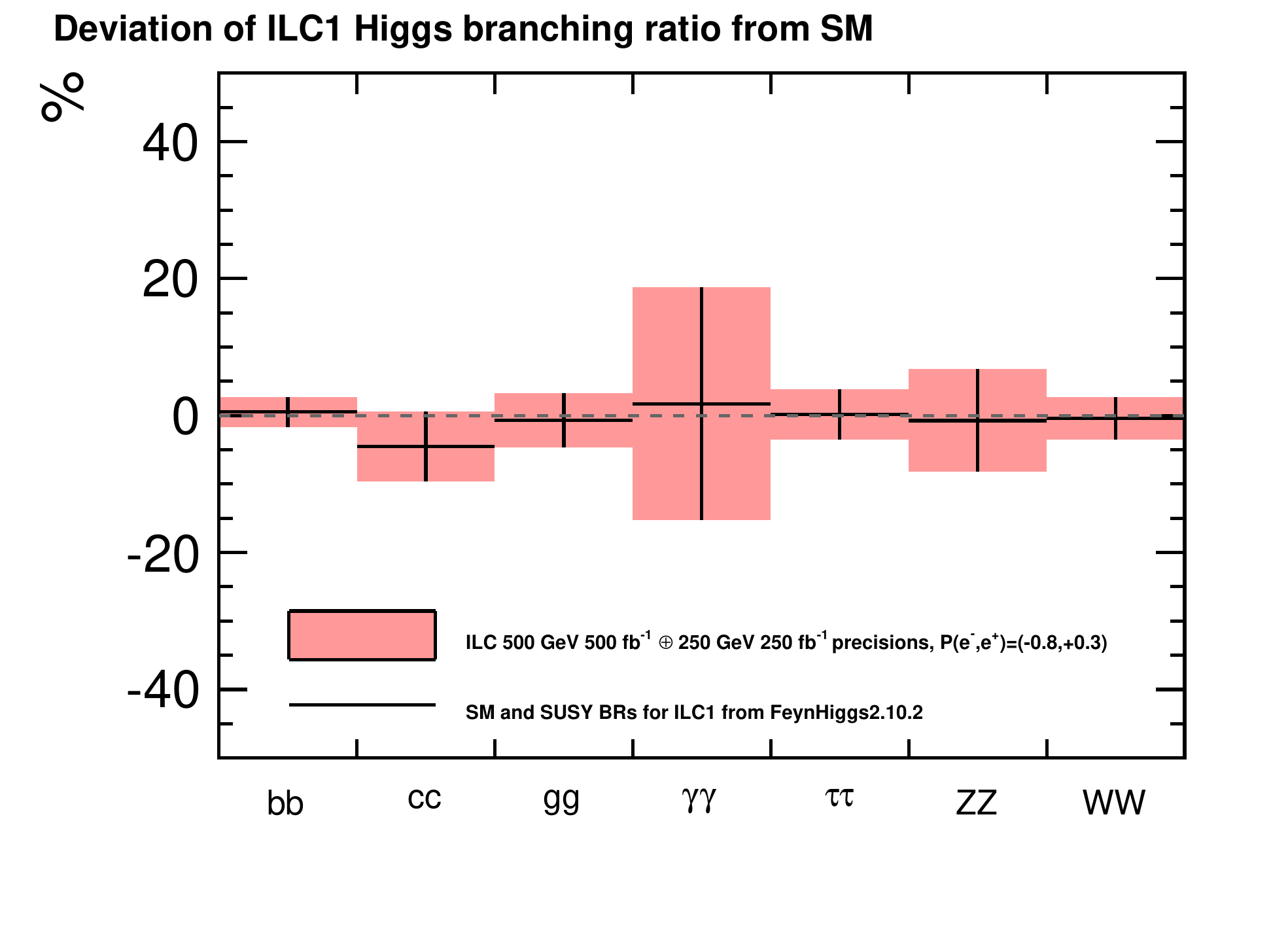}
\caption{ 
Left: Polarisation dependence of the production cross section for higgsino-like and wino-like charginos. 
Right: Expected Higgs branching ratio measurements for ILC1.
\label{fig:polxsec}}
\end{center}
\end{figure}

A further prediction of typical natural SUSY models is a nearly SM-like Higgs boson.
This is illustrated for the ILC1 benchmark in the right panel of Fig.~\ref{fig:polxsec},
where the shaded bands indicate the uncertainties of the model-independent coupling determination
after the initial stage of the ILC~\cite{Fujii:2015jha}. In spite of their SM-likeness, the
Higgs boson proporties will play an important role in the SUSY parameter determination.

The prospects for determining the weak-scale SUSY parameters from the measurements introduced in the
previous section have been evaluated using {\sc Fittino}~\cite{Bechtle:2004pc}. In addition to the ILC projections, we assumed a measurement of the gluino mass from the LHC with a precision of $10\%$. However this assumption is not critical for any other parameter than $M_3$. 
Parameters which do not enter the electroweakino sector at tree-level have been fixed to their true values here.
More general fits are in progress. Figure~\ref{fig:chi2} shows $\Delta \chi^2$ vs each of the fitted parameters.
\begin{figure}[tbp]
\begin{center}
\includegraphics[width=0.99\textwidth]{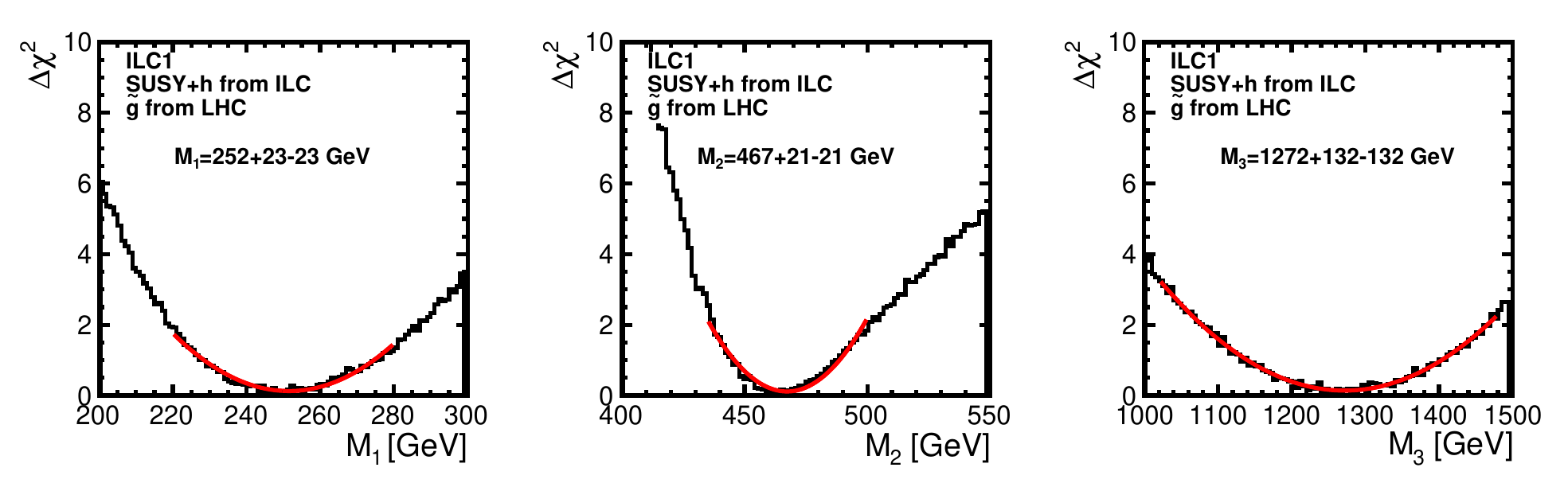}
\includegraphics[width=0.66\textwidth]{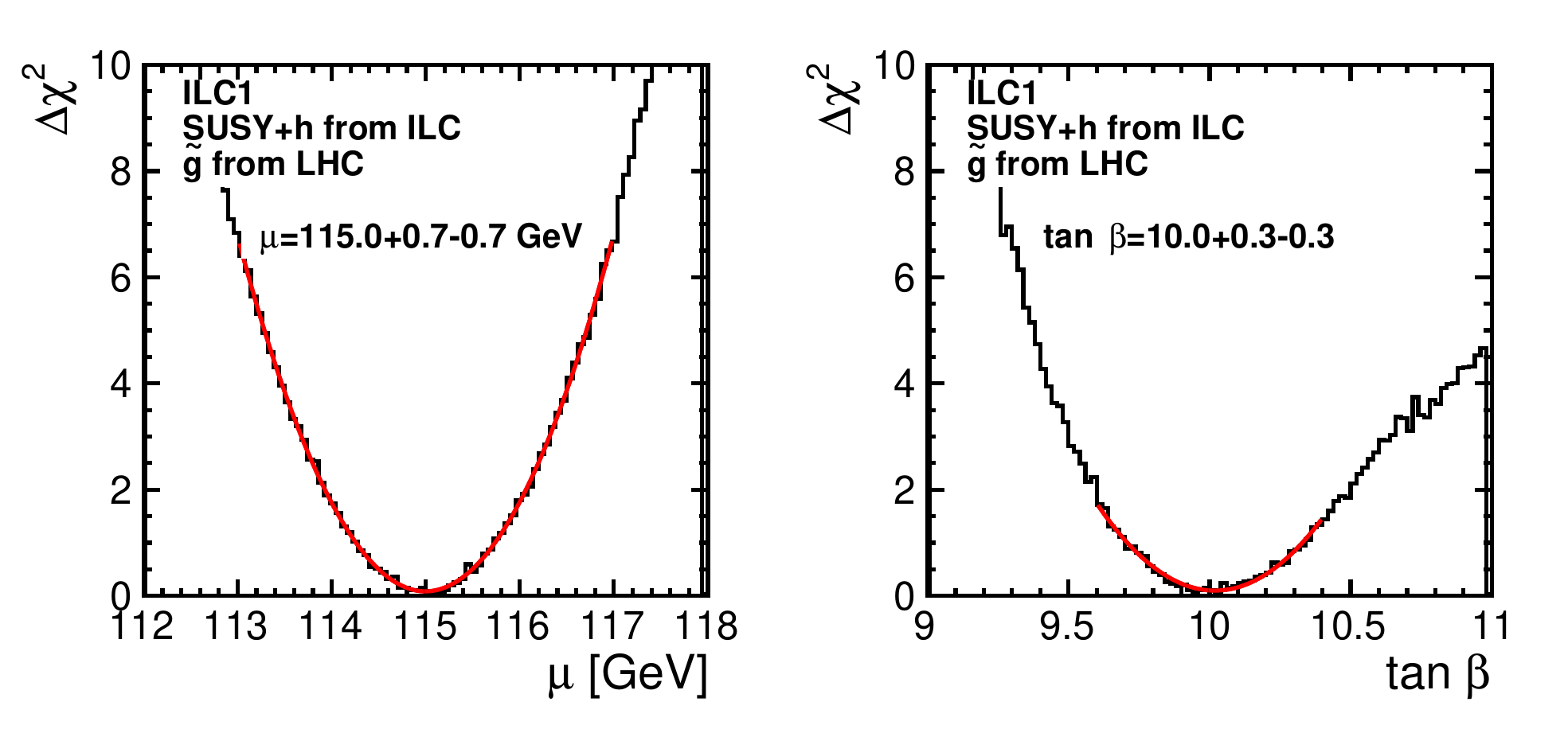}
\includegraphics[width=0.33\textwidth]{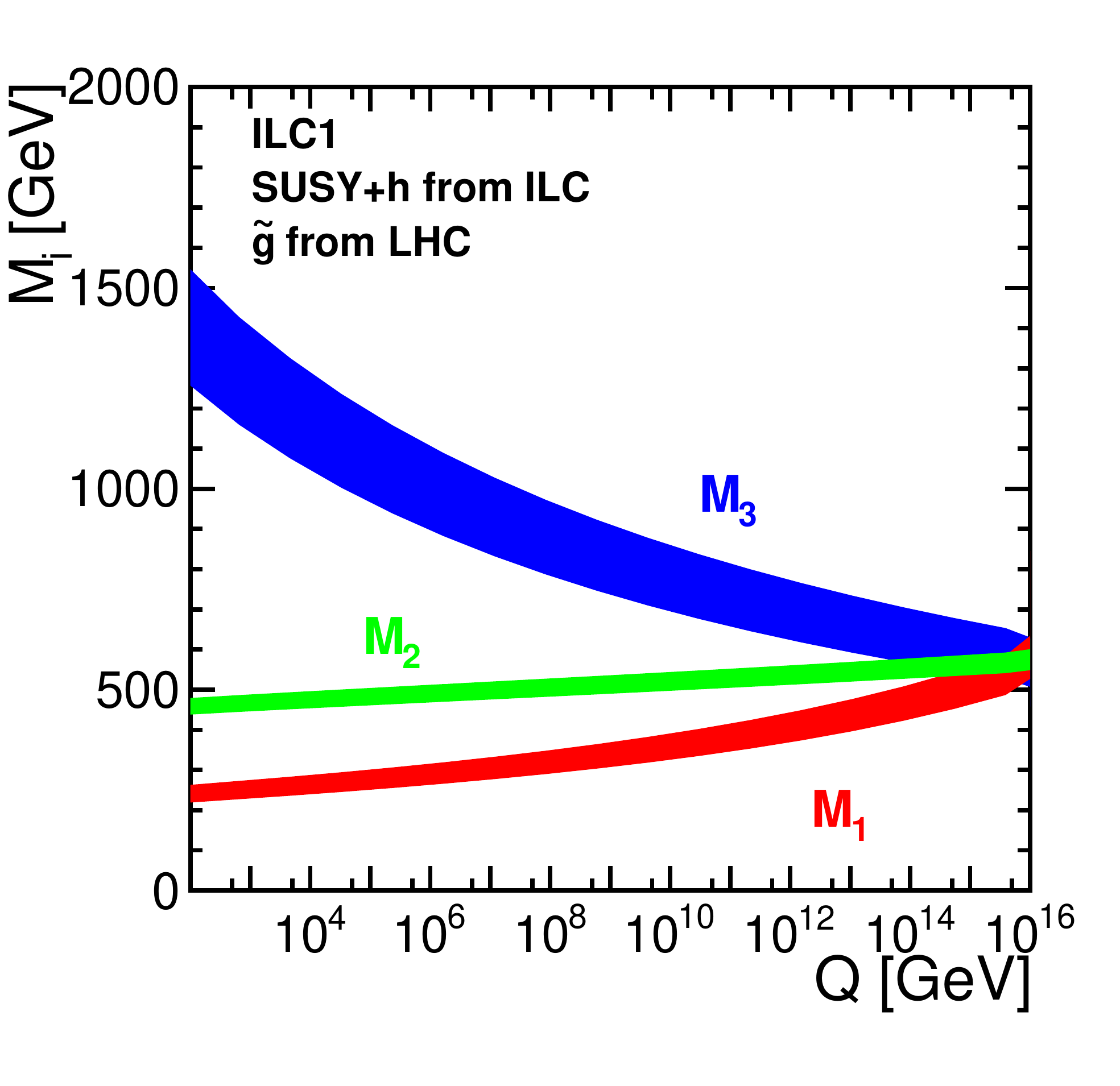}
\caption{$\Delta \chi^2$ vs $\mu$, $\tan\beta$, $M_1$, $M_2$, $M_3$. The central values and $1\sigma$ uncertainties are specified in the graphs. Lower right panel: running of the
fitted $M_1$, $M_2$ and $M_3$ values to higher scales indicates mass unification at the GUT scale.}
\label{fig:chi2}
\end{center}
\end{figure}

The resulting values and uncertainties for the weak-scale SUSY parameters can then be evolved
to the GUT scale using 2-loop RGEs, here using {\sc SPheno}~\cite{Porod:2003um}, as
shown in the lower right panel of Fig.~\ref{fig:chi2}. This way, the hypothesis of gaugino mass unification at the GUT scale (here due to the underlying NUHM2 benchmark) could be tested based on the joint capabilities of LHC and ILC. Should the gluino be unobservable at the LHC, still the energy scale $Q$ for unification of $M_1$ and $M_2$ could be tested. 
Assuming also that $M_3$ unifies with them at the GUT scale, the gluino mass could then be predicted, 
giving important information for the design of
the next generation hadron collider after the LHC.  

\section{Conclusions}

Motivated by the twin missives of simplicity and naturalness, then SUSY models with low $\mu$ and consequently
light higgsinos of mass $\sim 100-300$ GeV are highly favored. Such light higgsinos would be easily visible at ILC with
$\sqrt{s}>2m_{higgsino}$. Discovery of light higgsinos along with precision measurements of their properties would 
point to a natural origin for the EWSB sector and usher in a revolution in physics. 
This is indeed a powerful reason to promptly begin construction of ILC!

\end{document}